\documentclass[12pt]{article}

\usepackage{latexsym,amsmath,graphicx}

\DeclareFontFamily{OT1}{rsfs10}{}
\DeclareFontShape{OT1}{rsfs10}{m}{n}{ <-> rsfs10 }{}
\DeclareMathAlphabet{\mathscript}{OT1}{rsfs10}{m}{n}

\def\bcite{\@ifnextchar [{\@tempswatrue\@bcitex}{\@tempswafalse\@bcitex[]}}
\def\@bcitex[#1]#2{\if@filesw\immediate\write\@auxout{\string\citation{#2}}\fi
  \let\@bcitea\@empty
  \@bcite{\@for\@bciteb:=#2\do
    {\@bcitea\def\@bcitea{,\penalty\@m\ }%
     \def\@tempa##1##2\@nil{\edef\@bciteb{\if##1\space##2\else##1##2\fi}}%
     \expandafter\@tempa\@bciteb\@nil
     \@ifundefined{b@\@bciteb}{{\reset@font\bf ?}\@warning
       {Citation `\@bciteb' on page \thepage \space undefined}}%
     \hbox{\csname b@\@bciteb\endcsname}}}{#1}}
\def\@bcite#1#2{{#1\if@tempswa , #2\fi}}

\newcommand{\eq}[1]{equation~(\ref{#1})}
\newcommand{\eqs}[2]{equations~(\ref{#1}) and~(\ref{#2})}
\newcommand{\eqto}[2]{equations~(\ref{#1}) -~(\ref{#2})}

\newcommand{\ns}{\normalsize}

\newcommand{\bea}{\begin{eqnarray}}
\newcommand{\eea}{\end{eqnarray}}

\newcommand{\be}{\begin{equation}}
\newcommand{\ee}{\end{equation}}

\def\g{\gamma}

\def\f{\phi}
\def\vp{\varphi}

\def\l{\lambda}
\def\m{\mu}
\def\n{\nu}
\def\tn{\tilde \nu}

\def\r{\rho}

\begin{document}

%%%%%%%%%%%%%%%%%%%%%%%%%%%%%%%%%%%%%%%%%%%%%%%%%%%%%%%%%%%%%%%%%%%%%%

\begin{titlepage}

\title{
   \vspace{-4em}             
   \hfill{\ns BROWN-HET-1273}\\ 
   \hfill{\ns McGill 01-19}\\ 
   \hfill{\ns hep-th/0109165}
   \vfill
   {\LARGE Loitering Phase in Brane Gas Cosmology}}
%\author{
%   {\ns author1$^1$, author2$^2$
%      \setcounter{footnote}{1}\thanks{Lectures presented at the
%      Advanced School on Cosmology and Particle Physics, June 1998,
%      Peniscola, Spain.}~~
%     \setcounter{footnote}{2}\thanks{Supported in part by a Senior 
%          Alexander von Humboldt Award.}~~
%      and author3$^3$} \\
%   {\it\ns $^1$Department of Physics, Oxford University} \\[-0.5em]
%      {\it\ns 1 Keble Road, Oxford OX1 3NP, United Kingdom} \\   
%   {\it\ns $^2$Department of Physics, University of Pennsylvania} \\[-0.5em]
%      {\it\ns Philadelphia, PA 19104--6396, USA}\\
%   {\it\ns $^3$Department of Physics, Joseph Henry Laboratories,} \\[-0.5em]
%      {\it\ns Princeton University, Princeton, NJ 08544, USA}}
\author{
   {\ns Robert Brandenberger$^1$
      \setcounter{footnote}{1}\thanks{Address from 9/15/2001 - 3/15/2002:
Theory Division, CERN, CH-1211 Geneva, Switzerland}~~
     , Damien A. Easson$^{1,2}$ and Dagny Kimberly$^1$} \\
   {\it\ns $^1$Department of Physics, Brown University} \\[-0.5em]
      {\it\ns Providence, RI 02912, USA} \\[-0.5em]
      {\it\ns Email: rhb, easson@het.brown.edu}\\
   {\it\ns $^2$Physics Department, McGill University} \\[-0.5em]
      {\it\ns Montr\'eal, Qu\'ebec, Canada}}
\date{\ns \today}

\maketitle
\vspace{-1em}

\begin{abstract}
Brane Gas Cosmology (BGC) is an approach to M-theory cosmology
in which the initial state of the Universe is taken to be small, dense
and hot, with all fundamental degrees of freedom near thermal
equilibrium. Such a starting point is in close analogy with the 
Standard Big Bang (SBB) model. The topology of the Universe is 
assumed to be toroidal in all nine spatial dimensions
and is filled with a gas of p-branes.  The dynamics of winding 
modes allow, at most, three spatial dimensions 
to become large, thus
explaining the origin of our macroscopic $3+1$-dimensional Universe.
Here we conduct a detailed analysis of the loitering phase of BGC.  We do so
by including into the equations of motion that describe the dilaton gravity
background some new equations which determine the annihilation of string
winding modes into string loops.
Specific solutions are found within the model that exhibit loitering,
i.e. the Universe experiences a short phase of slow contraction
during which the Hubble radius grows larger than the physical extent of
the Universe.  As a result the brane problem (generalized domain wall problem)
in BGC is solved. 
The initial singularity and horizon
problems of the SBB scenario are  solved without relying on
an inflationary phase.

\end{abstract}
PACS numbers: 04.50+h; 98.80.Bp; 98.80.Cq.
\thispagestyle{empty} 

\end{titlepage}
%%%%%%%%%%%%%%%%%%%%%%%%%%%%%%%%%%%%%%%%%%%%%%%%%%%%%%%%%%%%%%%%%%%%%%%%%%%%

\section{Motivation and Introduction}

%%%%%%%%%%%%%%%%%%%%%%%%%%%%%%%%%%%%%%%%%%%%%%%%%%%%%%%%%%%%%%%%%%%%%%%%%%%%
The necessity to search for alternatives to the Standard Big Bang (SBB) 
scenario
is driven by a significant number of problems within the theory such as
the horizon, flatness, structure formation and cosmological constant problems.
Although inflationary models have managed to address many 
of these issues, all current formulations
remain incomplete.  In particular, the present models of inflation suffer from 
the fluctuation, super-Planck scale physics, initial singularity
and cosmological constant problems, as 
discussed in~\cite{Brandenberger:1999sw}.
Furthermore, several theorems have
appeared which show that de Sitter spacetime, the simplest example of an
inflationary Universe,  cannot be a classical solution to supergravity 
theories~\cite{Bautier:1997yp,Maldacena:2001mw}.
As supergravity is the low energy limit of $M$-theory, it does not seem 
clear that 
a pure de Sitter model of inflation will arise naturally within the theory.

$M$-theory is currently our best candidate for a quantum theory of gravity.  
As such, the theory
should provide the correct description of physics in regions of space with 
high energies and large curvature scales similar to those found
in the initial conditions of the Universe.  Therefore, it is
only natural to incorporate string and $M$-theory into models of 
cosmology.~\footnote{For a review of some string theory motivated 
cosmological models
see~\cite{Easson:2000mj}.  A recent proposal, not
covered in the review, is the ``Ekpyrotic" 
scenario introduced in~\cite{Khoury:2001wf}.}
It is our hope that $M$-theory will provide answers to the pending questions of
cosmology while preserving the triumphs of the SBB model ultimately leading 
to a complete description of the Universe.  

``Brane Gas Cosmology" (BGC) which is presented in~\cite{Alexander:2000xv} 
and based on the earlier 
work of~\cite{Brandenberger:1989aj} is one example of
a cosmological scenario motivated by $M$-theory.  The BGC model is 
relatively simple and starts out in close analogy with the Standard Big Bang 
scenario.  
According to~\cite{Alexander:2000xv}, the initial state of the
Universe is
small, dense, hot and with all fundamental degrees of freedom in 
approximate thermal 
equilibrium.~\footnote{Note that mathematically it is not possible for the 
Universe to 
be in thermal equilibrium as the FRW cosmological model does not possess a 
time-like
Killing vector field.  However, it is true that the Universe has been very 
nearly in thermal
equilibrium.  Obviously, the departures from equilibrium make things 
interesting!}
For simplicity, the background spatial geometry is assumed to be toroidal, 
and the Universe is
filled with a hot gas of $p$-branes, the fundamental objects appearing in
string theories.

These branes may wrap around the cycles of the torus
(winding modes), they can have a center-of-mass motion along the
cycles (momentum modes) or they may simply fluctuate in the bulk space 
(oscillatory modes).
By symmetry, we assume equal numbers of winding and anti-winding modes.
As the Universe tries to expand, the winding modes become heavy and halt the 
expansion~\cite{Tseytlin:1992xk}. Spatial dimensions can only
dynamically decompactify if the winding modes can disappear, and this is
only possible (for string winding modes) in $3+1$ 
dimensions~\cite{Brandenberger:1989aj}. Thus, BGC may provide
an explanation for the observed number of large spatial dimensions.
However, by causality at least one winding mode per Hubble volume will
be left behind, leading to the {\it brane problem} for 
BGC~\cite{Alexander:2000xv}, a problem analogous to the domain wall
problem of standard cosmology. 

The present paper provides a simple solution to the
brane problem: the winding modes will halt the expansion of the spatial
sections, and lead to a phase of slight contraction 
({\it loitering}~\cite{Sahni:1991ks,Feldman:1993ue})
during which the Hubble radius becomes larger than the spatial
sections and hence all remaining winding modes can annihilate in
the large $3+1$ dimensions. We supplement the equations for the
dilaton gravity background of BGC~\cite{Veneziano:1991ek,Tseytlin:1992xk} 
by equations which describe the annihilation of string winding modes
into string loops. A study of these equations demonstrates that solutions 
exist in which the winding modes force the Universe to contract for
a short time and enter a ``loitering" phase.
During the phase of contraction the number density of the
remaining winding modes increases and the winding
and anti-winding modes begin to annihilate.  
The winding branes appear as solitons (analogous to cosmic strings) 
in the bulk space.  The annihilation of winding and antiwinding modes 
(analogous to cosmic string intersections) leads to
the production of string loops which have the same equation of
state as cold matter.  

Brane Gas Cosmology is a simple nonsingular model which addresses some of 
the problems
of the SBB scenario, simultaneously providing a dynamical resolution to the 
dimensionality problem
of string theory.  In this respect it is very different from other attempts 
to incorporate $M$-theory into cosmology, such as ``brane world" scenarios 
in which our Universe is located on the world volume of a 3-brane.  In our 
opinion, brane
world models suffer from a lack of cosmological motivation.  
All current models  require that the extra dimensions be compactified by
hand.  Although this is a serious problem from a cosmological viewpoint 
it is rarely, if ever, discussed.

The organization of this paper is as follows.  
We begin with a brief review of the Brane Gas model in Section~\ref{BG}.
Our concrete starting point is presented, followed by a derivation of
the equation of state for a gas of branes and an analysis of the background
dynamics.   This is followed (in Section~\ref{LU}) by a discussion of the 
dilaton gravity equations in the presence of a brane gas, of the 
benefits of loitering, and of  attractor solutions.
In Section~\ref{LP}, we supplement the
system of equations with equations which describe the annihilation of
string winding modes into string loops, and based on this we provide a detailed
analysis of a loitering solution. 
We use both numerical and analytical methods to study the solutions.  

We conclude, in Section~\ref{CONC}, with a brief summary and a few 
conjectures concerning supersymmetry breaking,
the effective breaking of T-duality, and dilaton mass generation 
in the late Universe.
%%%%%%%%%%%%%%%%%%%%%%%%%%%%%%%%%%%%%%%%%%%%%%%%%%%%%%%%%%%%%%%%%%%%%%%%%%%%

\section{Brane Gases}\label{BG}

%%%%%%%%%%%%%%%%%%%%%%%%%%%%%%%%%%%%%%%%%%%%%%%%%%%%%%%%%%%%%%%%%%%%%%%%%%%%
The Brane Gas scenario of~\cite{Alexander:2000xv} is formulated within
the context of eleven-dimensional $M$-theory compactified on $S^1$.  This
leads to ten-dimensional, type II-A string theory, whose low-energy effective
action is that of supersymmetrized dilaton gravity.  

Since $M$-theory admits the graviton, 2-branes and 5-branes as
fundamental degrees of freedom, the compactification on $S^1$
leads to 0-branes, strings (1-branes), 2-branes, 4-branes,
5-branes, 6-branes and 8-branes in the ten-dimensional Universe.
The remaining nine spatial dimensions are assumed to be toroidal (of radius $R$).  
The Universe starts out hot, dense and near
thermal equilibrium.  It is filled with a gas of all the branes (wrapped and
unwrapped) which appear in the spectrum of the theory.
%%%%%%%%%%%%%%%%%%%%%%%%%%%%%%%%%%%%%%%%%%%%%%%%%%%%%%%%%%%%%%%%%%%%%%%%%%%%

\subsection{Brane Gas Equation of State:}\label{EOS}

%%%%%%%%%%%%%%%%%%%%%%%%%%%%%%%%%%%%%%%%%%%%%%%%%%%%%%%%%%%%%%%%%%%%%%%%%%%%
In this section we derive the equation of state of the brane gas
described above.  The gas consists of all the branes of spatial dimension $p$.
Contributions of the winding, momentum and
oscillatory modes are treated separately below.

The total action for the above model is the sum of the bulk effective
action and the action of all the branes in the gas.  The low-energy bulk 
effective action is
\begin{equation}\label{SBLK}
S_{bulk} = \frac{1}{2 \kappa^2}\int d^{10}x \sqrt{-G} e^{-2 \phi} 
\bigl[ R + 4 G^{\mu \nu} \nabla_\mu \phi \nabla_\nu \phi 
- \frac{1}{12} H_{\mu \nu \alpha} H^{\mu \nu \alpha} \bigr] 
\,,
\end{equation}
where $G$ is the determinant of the background metric $G_{\mu \nu}$,
$\phi$ is the dilaton, $H$ denotes the field strength corresponding to
the bulk antisymmetric tensor field $B_{\mu \nu}$, and $\kappa$ is
determined by the ten-dimensional Newton constant.

Fluctuations of each of the $p$-branes are described by the Dirac-Born-Infeld (DBI) 
action~\cite{Polchinski:1996na}
and are coupled to the ten-dimensional action via delta function sources.
The DBI action is
\begin{equation} \label{brane}
S_p \, = \, T_p \int d^{p + 1} 
\zeta e^{- \phi} \sqrt{- det(g_{mn} + b_{mn} + 2 \pi \alpha' F_{mn})}
\,,
\end{equation}
where $T_p$ is the tension of the brane, $g_{mn}$ is the induced metric on 
the brane, $b_{mn}$ is the induced antisymmetric tensor field, and $F_{mn}$ 
the field strength tensor of gauge fields $A_m$ living on the brane.  The
constant $\alpha' \sim l_{st}^2$ is given by the string length scale
$l_{st}$.

The induced metric on the brane $g_{mn}$, (with indices $m,n,...$ denoting 
spacetime dimensions 
parallel to the brane), is determined by the background metric $G_{\mu \nu}$ and by scalar 
fields $\phi_i$ (not to be confused with the dilaton $\phi$) living on the brane. 
The indices $i,j,...$ denote dimensions transverse to the brane and the $\phi_i$ describe 
the fluctuations of the brane in the transverse directions.  In the 
string frame, the tension of a $p$-brane, appearing in front of the 
action (\ref{brane}), is given by
\begin{equation}
T_p \, = \, {{\pi} \over {g_s}} (4 \pi^2 \alpha')^{-(p + 1)/2} \,,
\end{equation}
where $g_s$ is the string coupling constant.   We assume small string
coupling so that the fluctuations of the branes will be small.
We choose to work with conformal time $\eta$ in a background metric of the form
\begin{equation}\label{bkmet}
G_{\mu \nu} \, = \, a(\eta)^2 diag(-1, 1,\dots, 1) \, ,
\end{equation}
where $a(\eta)$ is the cosmological scale factor.

Assuming that the transverse fluctuations of the brane and the gauge 
fields on the brane are small, it is possible to expand the brane action as
\begin{eqnarray} \label{actexp}
S_{p} \, &=& \, T_p \int d^{p+1} \zeta a(\eta)^{p + 1} e^{-\phi} \nonumber \\
& & \times \, e^{{1 \over 2} 
tr log(1 + \partial_m \phi_i \partial_n \phi_i 
+ a(\eta)^{-2} 2 \pi \alpha' F_{mn})} \nonumber \\
&=& \, T_p \int d^{p + 1} \zeta a(\eta)^{p + 1} e^{- \phi} \\
& & \times \,(1 + {1 \over 2} (\partial_m \phi_i)^2 - \pi^2 
{\alpha'}^2 a^{-4} F_{mn}F^{mn}) \, . \nonumber
\end{eqnarray}
The first term inside the parentheses in the last line represents the
brane winding modes, the second term 
corresponds to the transverse fluctuations, and the third term relates to brane matter.  
In the low-energy limit, the transverse fluctuations of the 
brane are described by a free scalar field action, and the
longitudinal fluctuations are given by a Yang-Mills theory. The
induced equation of state which describes the second and third
terms has pressure $p \geq 0$.

We are now ready to compute the equation of state for the brane gases for
various $p$.  There are three types of modes that we will need to 
consider.  First, there are the winding modes.  The background space
is $T^9$, and hence a $p$-brane can wrap around any set
of $p$ toroidal directions.  These modes are related by T-duality to
the momentum modes corresponding to the center of mass motion of the
branes.  Finally, the modes corresponding to fluctuations of the branes
in the transverse directions are (in the low-energy limit) described
by the scalar fields on the brane, $\phi_i$.  There are also bulk matter
fields and brane matter fields.

Let us begin by considering the winding modes.  From equation
(\ref{actexp}), one can compute the equation of state for a winding
$p$-brane:
\begin{equation} \label{EOSwind}
\tilde{p} \, = \, w_p \rho \,\,\, {\rm with} \,\,\, w_p = - {p \over d} \,,
\end{equation}
where $d$ is the number of spatial dimensions, and $\tilde{p}$ and $\rho$ 
represent the pressure and energy density, respectively.

Both fluctuations of the branes and brane matter are described by free
scalar and gauge fields living on the brane.  These can be viewed as
particles in the directions transverse to the brane and thus the equation of
state is just that of ordinary matter:
\begin{equation} \label{EOSnw}
\tilde{p} \, = \, w \rho \,\,\, {\rm with} \,\,\, 0 \leq w \leq 1 \, .
\end{equation} 
{}From the action (\ref{actexp}) we see that the energy in the winding
modes will be
\begin{equation} \label{winden}
E_p(a) \, \sim \, T_p a(\eta)^p \, ,
\end{equation}
where the constant of proportionality is dependent on the number of
branes.  Note that the energy in the winding modes increases with the
expansion of the Universe in contrast to the energy of the brane fluctuations
and brane matter.
%%%%%%%%%%%%%%%%%%%%%%%%%%%%%%%%%%%%%%%%%%%%%%%%%%%%%%%%%%%%%%%%%%%%%%%%%%%%

\subsection{Background Equations of Motion:}\label{EOM}

%%%%%%%%%%%%%%%%%%%%%%%%%%%%%%%%%%%%%%%%%%%%%%%%%%%%%%%%%%%%%%%%%%%%%%%%%%%%
The background equations of motion are calculated from the variation of 
(\ref{SBLK}) with
the metric (\ref{bkmet}) and
were derived (in the absence of $H$)
in~\cite{Tseytlin:1992xk} (see also \cite{Veneziano:1991ek}).  
It is convenient to define
\begin{equation}
\lambda(t) \, = \, log (a(t)) \, ,
\end{equation}
and to use the ``shifted" dilaton
\begin{equation}
\varphi \, = \, 2 \phi - d \lambda \, ,
\end{equation}
which absorbs the space volume factor.  Assuming for simplicity the isotropic
case (all $a_i = a$ and therefore $\l_i = \l$) the
background equations are
\begin{eqnarray} \label{EOMback1}
- d \dot{\lambda}^2 + \dot{\varphi}^2 \, &=& \, e^{\varphi} E \\
\label{EOMback2}
\ddot{\lambda} - \dot{\varphi} \dot{\lambda} \, &=& \, {1 \over 2} e^{\varphi} P \\
\label{EOMback3}
\ddot{\varphi} - d \dot{\lambda}^2 \, &=& \, {1 \over 2} e^{\varphi} E \, ,
\end{eqnarray}
where $E$ and $P$ denote the total energy and pressure, respectively.  
Both sources $E$ and $P$ are made up from contributions of all the branes
in the gas.  We have already calculated these contributions in the
previous section: 
\begin{eqnarray}
E \, &=& \, \sum_p E_p^w + E^{nw} \nonumber \\
P \, &=& \, \sum_p w_p E_p^w + w E^{nw} \, ,
\end{eqnarray}
where the superscripts $w$ and $nw$ stand for the winding modes and
the non-winding modes, respectively.  Here the contributions of the
non-winding modes of all branes are combined into one term. The
constants $w_p$ and $w$ are given by (\ref{EOSwind}) and
(\ref{EOSnw}). Each $E_p^w$ is the sum of the energies of all of the
winding branes with dimension $p$.
%%%%%%%%%%%%%%%%%%%%%%%%%%%%%%%%%%%%%%%%%%%%%%%%%%%%%%%%%%%%%%%%%%%%%%%%%%%%

\subsection{Brane Gas Cosmology:}\label{BGC}

%%%%%%%%%%%%%%%%%%%%%%%%%%%%%%%%%%%%%%%%%%%%%%%%%%%%%%%%%%%%%%%%%%%%%%%%%%%%
We now turn our attention to the model of Brane Gas Cosmology
proposed in~\cite{Alexander:2000xv}. The first important aspect of BGC
is that it is free of the initial cosmological singularity of the
SBB model.  This can be explained by the T-duality symmetry of string theory 
as was first demonstrated 
in~\cite{Brandenberger:1989aj}.  As the Universe contracts, the T-duality
self-dual fixed point ($R=1$) is reached at some temperature $T$ less
than the limiting Hagedorn temperature.  As the background continues
to contract, the temperature \it decreases \rm according to the
T-duality equation for temperature,
\begin{equation} \label{tempdual}
T({1 \over R}) \, = \, T(R) \, .
\end{equation}
Thus, there is no physical singularity as $R$ approaches $0$.

The T-duality symmetry of the spectrum of states was analyzed
in \cite{Brandenberger:1989aj} in the absence of branes. 
With only fundamental string states,
T-duality interchanges the winding and momentum quantum numbers
of the same object. The action of T-duality on branes is more
complicated~\cite{Polchinski:rr}.  A T-duality parallel to a
$p$-brane produces a $(p-1)$-brane, while a T-duality perpendicular to a
$p$-brane yields a $(p+1)$-brane. Hence, after T-dualizing in all
$d$ spatial dimensions, a $p$-brane becomes a $(d-p)$-brane. The $p$
winding modes of the initial $p$-brane are heavy, and the $(d-p)$ transverse
momentum modes are light (considering as starting point $R >> 1$).
After T-dualizing, the $p$ transverse momentum modes of the $(d-p)$-brane
are heavy, and the $(d-p)$ winding modes are light (since now the
radius of the torus is $R << 1$). We conclude that the
spectrum of states respects T-duality in the presence
of branes.~\footnote{We thank S.~Alexander and F.~Quevedo for
key discussions on this point.}

Let us now examine the de-compactification mechanism which leads 
to three large spatial 
dimensions~\cite{Alexander:2000xv, Brandenberger:1989aj}.  
Recall, that
our starting point is a Universe with nine spatial dimensions all
equal and near the self-dual point, $R=1$.  The Universe is in thermal
equilibrium, and by symmetry we argue that there are equal numbers
of winding and anti-winding modes.

As the Universe begins to expand (symmetrically in all directions),
$\l$ increases and the total energy in the winding modes increases
according to \eq{winden}.  Note that the energy in the winding modes
of the branes with the largest value of $p$ increases the fastest.  Therefore
branes with the largest value of $p$ will have an important effect first.
As we have already explained, the winding modes will prevent further
expansion until they have annihilated.  

A simple counting argument demonstrates that the world-volumes of 
two $p$-branes will probably intersect in at most $2p+1$ spatial 
dimensions.~\footnote{For example, consider two
particles (0-branes) moving through a space of dimension $d$.  These particles
will definitely interact (assuming the space is periodic) if $d=1$, whereas they
probably will not find each other in a space with $d>1$.}  Clearly, in 
$d=9$ spatial dimensions the winding branes with $p=8,6,5$ and $4$ will
have no problems interacting and self-annihilating.  However, branes with
smaller values of $p$ will allow a hierarchy of dimensions to expand.
Since the energy of the branes with the largest value of $p$ is greatest,
the $2$-branes will first allow a $5$-dimensional torus to expand.
Within this $T^5$, the strings ($1$-branes) will allow a $T^3$
subspace to become large.  Hence, this mechanism provides a solution to
the dimensionality problem of string theory and may explain the origin of our
large $3+1$-dimensional Universe.~\footnote{By large we mean large compared 
to the string scale. Without inflation we are not able to solve the
entropy problem of standard cosmology, namely to produce a Universe large
enough to contain our present observed Universe.}

An important unsolved issue in Brane Gas Cosmology is the
stability of the radius of the compact dimensions to
inhomogeneities as a function of the three coordinates
$x_i$; $i = 1, 2, 3$ corresponding to the large spatial
dimensions. The separation in $x_i$ between the branes wrapping the 
small tori is increasing, and there appears to be no mechanism to keep the
``internal" dimensions from expanding (inhomogeneously in $x_i$)
between the branes. A simple but unsatisfactory solution, is to
invoke a non-perturbative effect similar to what is needed to
stabilize the dilaton at late times, namely to postulate
a potential which will stabilize the moduli 
uniformly in $x_i$ (after the winding modes around the $x_i$
directions have disappeared). Work on a possible solution within
the context of the framework presented here is in progress.~\footnote{We 
thank D. Lowe for discussions on this issue.}
%%%%%%%%%%%%%%%%%%%%%%%%%%%%%%%%%%%%%%%%%%%%%%%%%%%%%%%%%%%%%%%%%%%%%%%%%%%%

\section{A Loitering Universe}\label{LU}

%%%%%%%%%%%%%%%%%%%%%%%%%%%%%%%%%%%%%%%%%%%%%%%%%%%%%%%%%%%%%%%%%%%%%%%%%%%%
Despite the ability of winding modes to self-annihilate in a distinguished
number of dimensions, causality demands
that there is at least one winding mode per Hubble volume 
remaining~\cite{Kibble:1976sj}.
In our macroscopic four-dimensional spacetime, the wrapped branes with $p\ge 2$
will appear as domain walls.  The presence of such topological defects
would result in a cosmological
disaster, as even one wall per Hubble volume today will overclose the Universe
if the tension of the brane is larger than the electroweak 
scale.~\footnote{Even strings
have large tensions at low temperatures and will therefore contribute
too much to the energy density of the Universe.}
This ``domain wall problem"~\cite{Zeldovich:1974uw} is common in cosmological 
scenarios based on quantum field theories which admit domain wall
solutions.

As a solution of this ``brane problem" in BGC the authors
of~\cite{Alexander:2000xv}
suggested a phase of cosmological loitering.  If at some
stage in the evolution of the Universe the Hubble radius becomes larger than
the spatial extent of the Universe, there is no causal obstruction for all
winding modes to annihilate.  Loitering allows for previously 
causally disconnected regions of the Universe to communicate, and therefore
provides a solution to the horizon problem of the SBB model.  

Within the context of dilaton gravity,
cosmological solutions which exhibit a loitering phase appear rather
naturally due to the presence of winding modes.  To see this, we will
examine the phase space of solutions to the background \eqto{EOMback1}{EOMback3}.
The phase space of solutions for general $p$ was discussed in~\cite{Tseytlin:1992xk},
and a numerical plot of the full phase space is given below.

By defining $l = \dot\l$ and $f = \dot\vp$, the background EOM 
\eqto{EOMback1}{EOMback3} simplify to two first order differential
equations:
\begin{eqnarray}
\label{ldot}
\dot l \, &=& \, {p l^2 \over 2} + lf - \frac{pf^2}{2d} \,,\\
\label{fdot}
\dot f \, &=& \, {d l^2 \over 2} + {f^2 \over 2} \, .
\end{eqnarray}
Notice that for positive energy density $E$, \eq{EOMback1} implies  that $\dot\vp$ will never
change sign.  We are interested in studying the initial conditions with
$\dot\vp < 0$.  If $\dot\vp \not< 0$ the boosting effect of the dilaton on $\l$ will
invalidate the adiabatic approximation used in the derivation of the
EOM.  Furthermore, growing $\vp$ together with expanding $\l$ implies the
growth of the effective coupling $\exp(\f)$ in contradiction with a weak coupling
assumption~\cite{Tseytlin:1992xk}.

We will therefore consider solutions to the background EOM with initial
conditions corresponding to an expanding Universe $\dot \l>0$ with $\dot\vp < 0$.
Note that positivity of $E$ imposes another restriction (see \eq{EOMback1}):
\begin{equation}
\vert l \vert \, < \, {1 \over {\sqrt{d}}} \vert f \vert 
\,.
\end{equation}

Solutions with initial conditions described above are driven towards the $l=0$ line
in the $f$ vs. $l$ phase space at a finite value of $f$ (see Fig. 1).

\begin{figure}\label{phase1.eps}
\centering
%\begin{figure}[htbp]
\includegraphics[angle=270,width=4in]{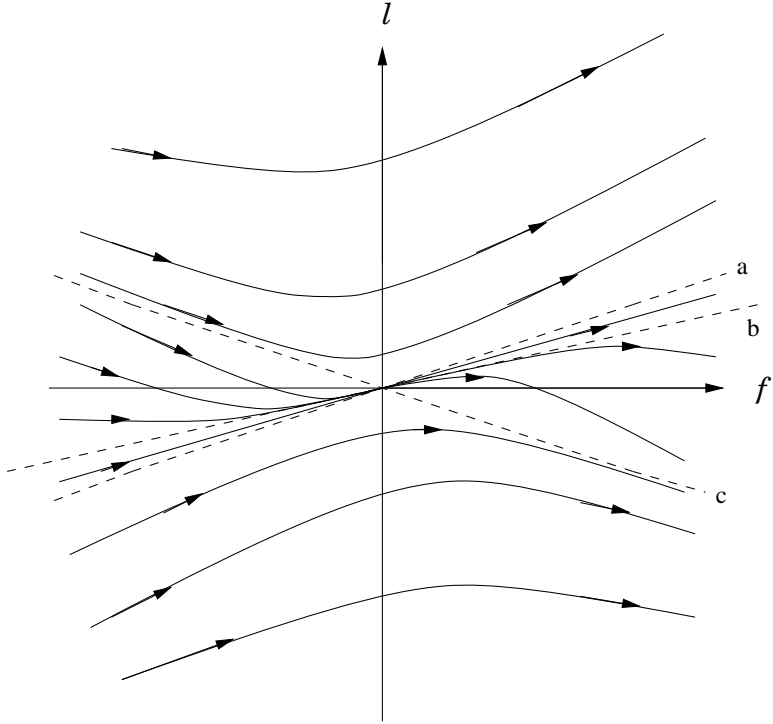}
\caption {Phase space trajectories of the solutions of the background equations (\ref{EOMback1} - \ref{EOMback3}) for the values $p = 2$ and $d = 9$. The energetically allowed region lies near the $l = 0$ axis between the special lines {\it a} and {\it c}, which are the lines given by $l/f = \pm 1 / \sqrt{d}$. The trajectory followed in the scenario investigated in this paper starts out in the upper left quadrant close to the special line {\it c} (corresponding to an expanding background), crosses the $l = 0$ axis at some finite value of $f$ (at this point entering a contracting phase), and then approaches the loitering point $(l, f) = (0, 0)$ along the phase space line {\it b} which corresponds to $l/f = p/d$.}
\end{figure}

There are three special lines in the phase space which correspond to straight
line trajectories and pass through the origin:
\begin{equation}
{{\dot l} \over {\dot f}} \, = 
\, {l \over f} \, = \, \pm {1 \over {\sqrt{d}}} \, , \, {p \over d} \, .
\end{equation}
Notice that the line $l/f = - 1/\sqrt{d}$ is a repeller and solutions which
start out in the energetically allowed region are pushed away from the
line, cross the $f$-axis and then approach the attractor line  
$l/f = p/d$ to the origin.  Near the $l=0$ line, \eqs{ldot}{fdot} may
be approximated by
\begin{equation}
{\dot l} \, \simeq \, - {p \over {2d}} f^2 \, , \, {\dot f} \, \simeq \, {{f^2} \over 2} \, .
\end{equation}
When a solution crosses the $f$-axis, $l$ changes sign.  This means that
the Universe begins to contract.  Since $f \rightarrow 0$ only as 
$t \rightarrow \infty$ the solution never crosses the $l$-axis.

Note that this analysis assumes that the winding modes are not decaying 
into loops.  
The above provides an accurate description of the early Universe, before 
winding modes
have self-annihilated.  We will examine the case of loop production and the
late time evolution of the Universe in Section~\ref{LP}.
%%%%%%%%%%%%%%%%%%%%%%%%%%%%%%%%%%%%%%%%%%%%%%%%%%%%%%%%%%%%%%%%%%%%%%%%%%%%

\section{Unwinding and Loop Production}\label{LP}

%%%%%%%%%%%%%%%%%%%%%%%%%%%%%%%%%%%%%%%%%%%%%%%%%%%%%%%%%%%%%%%%%%%%%%%%%%%%
We now wish to extend the analysis presented in Section~\ref{LU} in order to
study the late time behavior of the Universe,
i.e. to include the effects of winding mode annihilation and loop production.

Recall that after the winding modes have annihilated, a three-dimensional
subspace
will grow large.  In what follows we will therefore take $d=3$.
The strings in the theory are the last branes to unwind which implies at late times that
we should consider the case of $p=1$.  When the winding strings self-annihilate
they create loops in the $3+1$-dimensional universe.

We now set up the equations describing the unwinding and corresponding
loop production. They are analogous to the corresponding equations for
cosmic strings in an expanding Universe. First, note that the energy
density $\rho_w$ in winding strings can be expressed in terms of the
string tension $\mu$ and the number ${\tilde \nu}(t)$ of winding
modes per ``Hubble" volume $t^3$ as 
\begin{equation} \label{def1}
\r_w(t) \, = \, \m \tn(t) \, t^{-2} \, .
\end{equation}
Since loops are produced by the intersection of two winding strings,
the rate of loop production is proportional to ${\tilde \nu}^2$:
\begin{equation}
\frac{dn(t)}{dt}\, = \, c {\tn(t)}^2 \, t^{-4} \,, 
\end{equation}
where $n(t)$ is the number density of loops and $c$ is a 
proportionality constant expected to be of the order $1$. The
energy density in the winding modes decreases both due to the
expansion of space and due to the decay into loops:
\begin{equation} \label{energy1}
\frac{d\r_w(t)}{dt} + 2 l \, \r_w(t) 
        \, = \, -c'\m t \frac{dn(t)}{dt} \, = \, -cc'\m{\tn(t)}^2 \,
t^{-3} \,,
\end{equation}
where $c^{'}$ is a constant which relates the mean
radius $R = c^{'} t$ of a string loop to its length. 
Without loop production ($c = 0$), the energy density 
$\rho_w$ redshifts corresponding to the equation of state
$p = - {1 \over 3} \rho$. This explains the coefficient of
the Hubble damping term in (\ref{energy1}).~\footnote{For
more on strings in an expanding Universe see, e.g.~\cite{Vilenkin:1994}.}
Inserting \eq{def1} into the energy conservation \eq{energy1}, we obtain
an equation for $\tn(t)$:
\begin{equation}\label{eomn}
\frac{d\tn(t)}{dt}\, = \, 2 \tn (t^{-1} - l) - cc't^{-1} \tn^2
\,.
\end{equation}

In addition to $\r_w(t)$, we will also require information about the 
energy density in loops, $\r_l(t)$.
The energy density in loops obeys the conservation equation
\begin{equation}\label{julia}
\frac{d\r_l(t)}{dt} + 3 l \, \r_l(t) 
        \, = \,  cc'\m{\tn(t)}^2 \, t^{-3}
\,.
\end{equation}
As in \eq{energy1}, the second term on the left hand side of
the equation represents the decrease in the density due to
Hubble expansion, with the coefficient reflecting the equation
of state $p = 0$ of a gas of static loops, and the term on the
right hand side representing the energy transfer from winding
modes to loops. Without loop production, $\rho_l(t)$ would
scale as
\begin{equation}\label{rhol}
\r_l(t) \, = \, g(t) e^{-3(\l(t) - \l_0)} 
\, ,
\end{equation}
with $g(t)$ constant. Here $\l_0 = \l(t_0)$, where $t_0$ is some initial
time, and $g(t)$ is a function which obeys the equation
\begin{equation}\label{eomg}
\frac{dg(t)}{dt} \, = \, cc'\m t^{-3} \tn^2 e^{3(\l(t) - \l_0)}
\,.
\end{equation}

Using the expressions for $\r_w(t)$ and $\r_l(t)$ as sources for the energy
density $E$ and pressure $P$ in \eqto{EOMback1}{EOMback3} we can obtain
background equations analogous to \eqs{ldot}{fdot}:
\begin{eqnarray}
\label{eoml}
\dot l \, &=& lf +\frac{1}{2} l^2 - \frac{1}{6} f^2 + \frac{1}{6} g e^{\vp + 3\l_0}\,,\\
\label{eomf}
\dot f \, &=& \, \frac{1}{2} f^2 + \frac{3}{2} l^2 \, .
\end{eqnarray}
(Recall that $p=1$ and $d=3$ in the above equations.)  

Equations (\ref{eomn}), (\ref{eomg}), (\ref{eoml}) and (\ref{eomf}), along with
the equations $l = \dot\l$ and $f = \dot \vp$ provide six first-order differential
equations which fully describe
the Universe during the process of loop production.
These provide us with the
initial conditions required for a numerical analysis.
Note that the $c=0$ case corresponds to no loop production and the background equations
reduce to the previous \eqs{ldot}{fdot}.  

Figure 2 demonstrates the behavior of a typical numerical solution to the
EOM having initial conditions in the energetically allowed region of the
phase space, taking into account the effects of loop production.  Recall 
that when no loops are produced (see Fig. 1) the 
solutions of interest cross the $l=0$ line only once and approach
the origin of the phase space as $t \rightarrow \infty$.  When the decay
of the winding modes is taken into account, the solutions are pushed back
over the $l=0$ line as in Fig. 2.

\begin{figure}
\centering
%\begin{figure}[htbp]\label{phase2.eps}
\includegraphics[angle=0,width=3in]{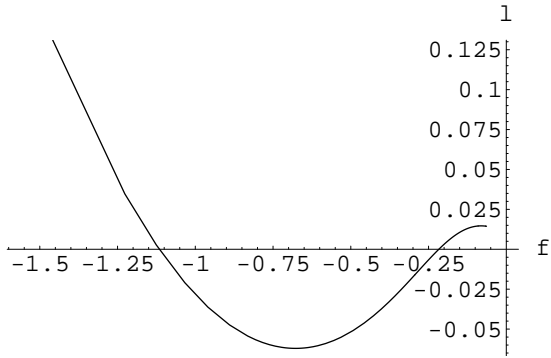}
\caption {A solution of the background equations (\ref{eoml}) and (\ref{eomf}) including
the effects of loop production. 
%for the values $p = 2$ and $d = 9$\footnote{Note
%that we are interested in the $p=1$, $d=3$ case at late times.  However
%the general behavior of the phase space does not change.}. 
This depicts a typical solution which starts in the energetically allowed region 
of the phase space. The solution crosses the $l = 0$ axis at some finite value of 
$f$ (at which point the Universe enters a contracting phase), 
and then crosses the
$l=0$ line a second time when the winding modes have fully annihilated.  At this
point the Universe begins to expand, is matter dominated and the dilaton is 
assumed to become massive.}
\end{figure}

In more detail, the dynamics of our loitering solution is
depicted in Figures 3, 4, 5 and 6. Figure 3 shows the time evolution
of the Hubble expansion rate $H = l$. Note that since we have set
Newton's constant $G = 1$ in our background equations, time is
measured in Planck time units. 
\begin{figure}\label{tvsl.eps}
\centering
%\begin{figure}[htbp]\label{tvsl.eps}
\includegraphics[angle=0,width=3in]{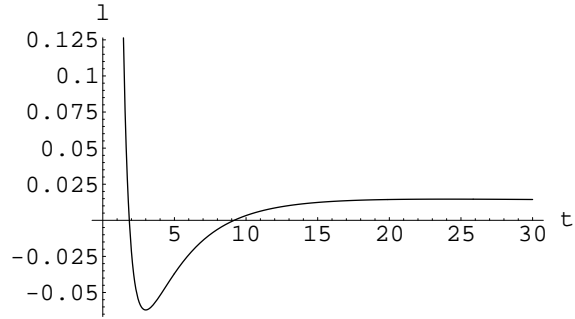}
\caption {The time evolution of $H = l$.  The loitering phase begins
when $l(t)$ crosses the $l =0$ line for the first time and ends when
$l(t)$ crosses back over the $l=0$ line.}
\end{figure}
By comparing with the value of
$a(t)$ from Fig. 4, we see
that
\begin{equation} \label{ineq}
H^{-1}(t) \, \gg \, a(t) 
\end{equation}
during the loitering phase.
\begin{figure}\label{tvsa.eps}
\centering
%\begin{figure}[htbp]
\includegraphics[angle=0,width=3in]{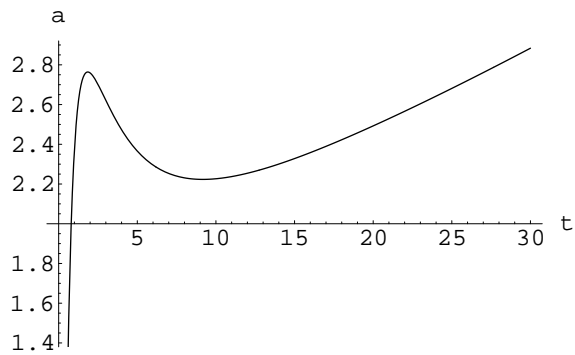}
\caption {The time evolution of the  scale factor $a$.  By comparing
this plot with Fig. 3 we see that the loitering phase 
lasts long enough to allow all winding modes to self-annihilate in the 
large three-dimensional Universe.}
\end{figure}
Keeping in mind that the initial
spatial size of the tori is Planck scale, it follows immediately
from \eq{ineq} and from the time duration of the loitering phase
that loitering lasts sufficiently long to allow causal
communication over the entire spatial section. This is 
reflected in Fig. 5 which shows that the
winding modes completely annihilate by the end of the
loitering phase, after which $g(t)$ tends to a constant (Fig 6).
\begin{figure}\label{tvsn.eps}
\centering
%\begin{figure}[htbp]
\includegraphics[angle=0,width=3in]{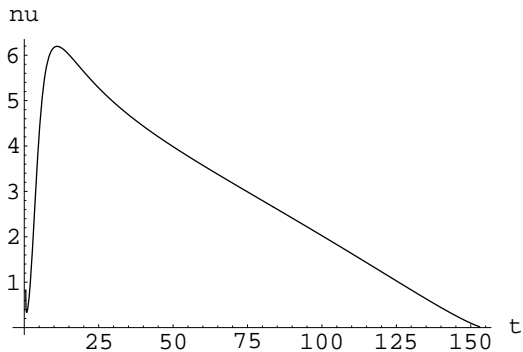}
\caption {Time evolution of $\tilde\n$. Initially, $\tilde\n$
  increases as
the Universe contracts.  The winding modes begin to self-annihilate
($\tilde\n$ decreases) and eventually vanish ($\tilde\n
\rightarrow 0$).}
\end{figure}
\begin{figure}\label{tvsg.eps}
\centering
%\begin{figure}[htbp]\label{tvsg.eps}
\includegraphics[angle=0,width=3in]{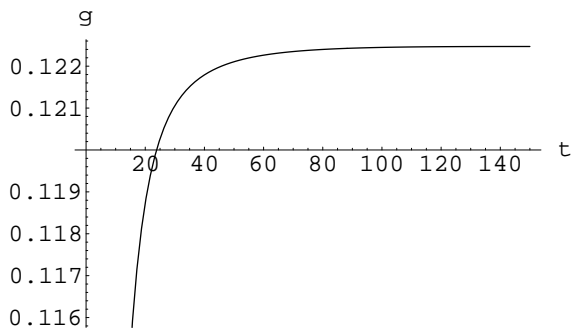}
\caption {Time evolution of $g$.  Note that $g$ goes to a constant
  when
$\tilde\n$ goes to zero.}
\end{figure}

Our modified picture is as follows: the Universe begins to expand until 
the winding modes become too massive and force the expansion to stop and contraction to begin.
This corresponds to the solution in Fig. 2 crossing over the $l=0$ axis and
signals the beginning of the loitering phase.  The loitering phase ends when the 
solution is pushed back over the $l=0$ line due to the decay of the
winding modes.  From this point on the Universe begins to expand again.

Soon after the solutions cross the $l=0$ line for the second time, our equations become singular.
Soon after the winding modes vanish ($\tn(t) \rightarrow 0$),
all the fields $l$, $f$, $\l$ and $\vp$ in our equations of motion blow up.  
This singularity does not concern us however since it can be eliminated by 
the introduction of a simple
potential $V(\f)$ used to freeze the dilaton at the moment loop production
is exhausted.  The massless dilaton does not appear in nature and therefore
such a potential is required in any string theory-motivated cosmological model at late times.
The precise mechanism responsible for dilaton mass generation is unknown,
although it is often suspected that this mechanism will coincide with the breaking
of supersymmetry.  From this analysis we are lead to the conjecture that
dilaton mass generation may coincide with the elimination of winding modes.
We will comment more on this below.

For the time being, let us assume that the dilaton has frozen at the value
$C_\f = \f(t_{freeze})$ and therefore $\dot\f = \ddot\f = 0$.
We will also assume that this occurs when the winding modes have vanished,
$\tn(t) \rightarrow 0$ and hence the number of 
loops has reached a constant so that $g(t) \rightarrow C_g$.  Now the EOM simplify
greatly.  By fixing the dilaton in \eqs{eoml}{eomf} we can derive an 
equation for the scale factor (after shifting back to the true dilaton $\f$):
\begin{eqnarray}\label{matdom}
\dot a - C_\g a^{-\frac{1}{2}} = 0
\,,
\end{eqnarray}
where $C_\g$ is a constant given by
\begin{eqnarray}
C_\g = \sqrt{\frac{ C_g }{12}} \, e^{ C_\f + \frac{3}{2}\l_0}
\,.
\end{eqnarray}
The most general solution to~\eq{matdom} is
\begin{eqnarray}\label{matsol}
a(t) \, = \, \left(\frac{3 C_\g}{2}\right)^{(\frac{2}{3})} \, 
         (t^2 - 2 C t + C^2)^{(\frac{1}{3})}
\,,
\end{eqnarray}
where $C$ is an integration constant.  For the value $C=0$ or for large 
values of $t$ (late times), the scale factor grows as
\begin{equation}\label{mdom}
a(t) \sim t^{\frac{2}{3}}
\,,
\end{equation}
which is exactly the behavior for a matter dominated Universe.

Our interpretation is that the winding modes look like solitons in the
$3+1$-dimensional Universe.  The self-annihilation of these winding modes
corresponds to the creation of matter in the Universe and the
scale factor evolves appropriately. In our equations, the loops are
modelled as static. In reality, the loops will oscillate and decay
by emitting (mostly) gravitational radiation, thus producing a radiation
dominated Universe.

Let us return to the issue of SUSY breaking and dilaton mass generation.
One thing which appears inevitable within the context of this model 
is the spontaneous ``breaking" of T-duality in the large four-dimensional 
Universe.  This is most easily understood once all of the winding 
modes have self-annihilated since it is impossible
to create new ones.  It would cost too much energy for a brane to wrap around
the large dimensions. Thus, the state of the system is not symmetric
under T-duality, and in the absence of string winding modes and for
fixed dilaton, our background equations  
reduce to those of Einstein's General Relativity 
which do not exhibit the $R \leftrightarrow 1/R$ symmetry of string theory.

It is also interesting to note that there seems to be a relation 
between the amount of supersymmetry in a theory and the presence of
T-duality.  Using a specific example in~\cite{Aspinwall:1999ii}, 
Aspinwall and Plesser show that
T-duality can be broken by nonperturbative effects in string coupling. 
Furthermore, a holonomy argument is given to show that T-dualities
should only be expected when large amounts of supersymmetry are present.
It seems very likely that the dynamics in the BGC scenario
will cause SUSY to break.  
This result seems to be in agreement with the possibility of dilaton 
mediated SUSY breaking occurring simultaneously with the breaking of T-duality. 

Considering the above evidence we are inclined to hypothesize about the 
possible relations
between supersymmetry breaking and the breaking of T-duality, as well as
dilaton mass generation and the vanishing of winding modes.
%%%%%%%%%%%%%%%%%%%%%%%%%%%%%%%%%%%%%%%%%%%%%%%%%%%%%%%%%%%%%%%%%%%%%%%%%%%%

\section{Conclusions and Speculations}\label{CONC}

%%%%%%%%%%%%%%%%%%%%%%%%%%%%%%%%%%%%%%%%%%%%%%%%%%%%%%%%%%%%%%%%%%%%%%%%%%%%
In this paper we have conducted a detailed analysis of a loitering
phase in the model of Brane Gas Cosmology presented in~\cite{Alexander:2000xv}.
The concrete starting point of BGC is $M$-theory compactified on $S^1$ which
gives ten-dimensional, type-IIA string theory.  We assume toroidal topology
in all nine spatial dimensions.  The initial conditions for the Universe
include a small, hot, dense gas of the $p$-branes in the theory.  These
fundamental degrees of freedom are assumed to be in thermal equilibrium.
T-duality ensures that the initial singularity of the SBB model is not present in
this scenario.

We compute the equation of state for the brane gas system and the background
equations of motion.  We study the solutions which initiate in the energetically
allowed region of the phase space.
The Universe tries to expand until winding modes
force the expansion to stop and a phase of slow contraction (loitering) to begin.
Loitering
provides a solution to the brane problem
first discussed in~\cite{Alexander:2000xv}.  It also provides a solution to
the horizon problem of the SBB model without requiring an inflationary
phase.~\footnote{Here
we do not address the structure formation problem
or the flatness problem of the SBB model.  Both of these are solved
by an inflationary phase. It is likely that the brane gas scenario will
require something like
inflation in order to produce a Universe which is large enough to
contain the known Hubble radius.}

The counting
argument of~\cite{Alexander:2000xv} demonstrates that winding modes will allow 
a hierarchy of dimensions in the $T^9$ to grow large.  When the string
winding modes self-annihilate we are left with a large $T^3$  
subspace, simultaneously
explaining the origin of our $3+1$-dimensional Universe and solving
the dimensionality problem of string theory.  

Branes wrapped around the cycles of the torus appear as solitons 
in the early Universe. They are
topological defects (domain walls for $p \ge 2$).  When the winding
modes and anti-winding modes self-annihilate, matter is 
produced and the Universe begins to expand again.
We hypothesize that winding mode annihilation corresponds to dilaton mass
generation.  We also believe there may be a relation between SUSY breaking and
the breaking of T-duality, although we cannot provide any direct evidence for this.
Once winding states have vanished, we cannot
map momentum modes into winding modes via T-duality.  The breaking
of T-duality requires further study.

Particle phenomenology demands compactification on manifold
with nontrivial holonomy such as Calabi-Yau three-folds
if the four-dimensional low energy effective theory is to have $N=1$ 
supersymmetry. The initial steps towards
generalizing the brane gas model to manifolds of
nontrivial homology are discussed in~\cite{Easson:2001fy}.
However, in the context of early
Universe cosmology it is not reasonable to
require $N=1$ supersymmetry. Toroidal backgrounds, such
as the one considered here, are
compatible with maximal supersymmetry.

Brane Gas Cosmology provides a method of incorporating
string and $M$-theory into cosmology which is an alternative 
to popular ``brane world" scenarios.
In our opinion, BGC has the advantage over brane world scenarios in that its 
foundations are analogous to those of the Standard Big Bang model.  
In the BGC model 
the Universe starts out small, hot and dense, with no
initial singularity.   
All the current versions of brane
world scenarios embedded in string theories rely on the compactification of the
extra dimensions by hand.  In our opinion this is a considerable problem which
is often overlooked.  A dynamical mechanism in BGC leads naturally to four 
large space-time dimensions. 

\vspace{0.4cm}
\subsection*{Acknowledgments:}
We are thankful to S. Alexander, D. Lowe and F. Quevedo for important
discussions. 
RB was supported in part by the U.S. Department of Energy under
Contract DE-FG02-91ER40688, TASK A.
DE was supported in part by the 
U.S. Department of Education under the GAANN program.
%%%%%%%%%%%%%%%%%%%%%%%%%%%%%%%%%%%%%%%%%%%%%%%%%%%%%%%%%%%%%%%%%%%%%%%%%%%%

%%%%%%%%%%%%%%%%%%%%%%%%%%%%%%%%%%%%%%%%%%%%%%%%%%%%%%%%%%%%%%%%%%%%%%%%%%%%

\end{document}